\documentclass[draft,tightenlines,nofootinbib,preprint,aps,eqsecnum,amsmath,amssymb]{revtex4}

\newcommand{\beq}{\begin{equation}}
\newcommand{\eeq}{\end{equation}}
\newcommand{\bea}{\begin{eqnarray}}
\newcommand{\eea}{\end{eqnarray}}

\newcommand{\sgn}{\epsilon}

\begin{document}

\title{Relativistic Entanglement from Relativistic Quantum Mechanics in the Rest-Frame
Instant Form of Dynamics.}

\medskip

\author{Luca Lusanna}

\affiliation{ Sezione INFN di Firenze\\ Polo Scientifico\\ Via Sansone 1\\
50019 Sesto Fiorentino (FI), Italy\\ Phone: 0039-055-4572334\\
FAX: 0039-055-4572364\\ E-mail: lusanna@fi.infn.it}

\begin{abstract}

After a review of the problems induced by the Lorentz signature of
Minkowski space-time, like the need of a clock synchronization
convention for the definition of 3-space and the complexity of the
notion of relativistic center of mass, there is the introduction of
a new formulation of relativistic quantum mechanics compatible with
the theory of relativistic bound states. In it the zeroth postulate
of non-relativistic quantum mechanics is not valid and the physics
is described in the rest frame by a Hilbert space containing only
relative variables. The non-locality of the Poincare' generators
imply a kinematical non-locality and non-separability influencing
the theory of relativistic entanglement and not connected with the
standard quantum non-locality.

\bigskip

Talk at the {\it Fifth Int. Workshop DICE2010 Space-Time-Matter},
Castiglioncello, September 13-17, 2010

\end{abstract}

\maketitle

\vfill\eject

Atomic physics is an approximation to QED, in which the atoms are
described as non-relativistic particles in quantum mechanics (QM)
with a coupling to the electro-magnetic field of order  $1/c$. For
all the applications in which the energies involved do not cross the
threshold of pair production, this description with a fixed number
of particles is enough. Therefore atomic physics and the theory of
entanglement are formulated in the absolute Euclidean 3-space and
use Newton absolute time, namely they are formulated in Galilei
space-time. The main drawback is that, due to the coupling to the
electro-magnetic field there is not a realization of the kinematical
Galilei group  connecting non-relativistic inertial frames. On the
other hand, if we want to arrive at an understanding of relativistic
entanglement, we must reformulate the theory in Minkowski space-time
with a well defined realization of the kinematical Poincare' group
connecting relativistic inertial frames. This would lead to {\it
relativistic atomic physics} as the quantization of a fixed number
of classical relativistic charged scalar (or spinning) particles
interacting with the classical electro-magnetic field.\medskip

In this framework it is possible to take into account the non-local
aspects of the Lorentz signature of space-time and of the Poincare'
group. Some of the implications are: i) the non-existence of a
unique notion of relativistic center of mass; ii) the propagation of
rays of light along the light-cone as implied by Maxwell equations.
Instead in the non-relativistic theory used for experiments testing
entanglement the world-lines of photons do not exist and only their
polarization is described by means of a two-dimensional Hilbert
space. While for experiments on Earth this is enough, teleportation
between the Earth and the Space Station \cite{1} will require a
control on the world-lines of light rays. See Refs.\cite{2,3} for
the existing attempt to formulate a relativistic theory of
entanglement.\medskip

Till now all the attempts to define relativistic QM employ the
so-called {\it zeroth postulate of QM} (see Zurek in Ref.\cite{4}).
According to it a composite system of two spatially separated
subsystems is described by the {\it tensor product} of the Hilbert
spaces of the subsystems. The notation ${\cal H} = {\cal H}_1
\otimes {\cal H}_2 = {\cal H}_{com} \otimes {\cal H}_{rel}$ means
that the quantum 2-body isolated system can be imagined to be
constituted either by the two single particle subsystems with masses
$m_1$ and $m_2$ or as the tensor product of a decoupled
center-of-mass particle of mass $m = m_1 + m_2$ carrying an internal
space with an internal relative motion of reduced mass $\mu = m_1\,
m_2/m$  \footnote{The zeroth postulate, i.e. ${\cal H} = {\cal H}_1
\otimes {\cal H}_2$, is based on a notion of {\it separability}
(Einstein's one) independent from the Galilei group, which instead
is at the basis of the decomposition ${\cal H} = {\cal H}_{com}
\otimes {\cal H}_{rel}$ emphasizing that the center-of-mass momentum
is a constant of motion for an isolated system so that one can do
the separation of variables in the Schroedinger equation.}. The two
descriptions are connected by a unitary transformation and
correspond to different choices of bases in ${\cal H}$. We will see
that the zeroth postulate does not hold at the relativistic level,
where the basic conceptual problem is the absence of an intrinsic
notion of 3-space without which we cannot formulate a well posed
Cauchy problem for Maxwell equations and we loose predictability.
This is the problem of clock synchronization, whose clarification
came from the attempts to get a consistent description of
relativistic bound states.

\medskip

Moreover the solution must be such that the transition from a
simultaneity convention to another one has to be formulated as a
gauge transformation (so that physical results are not influenced by
the convention, which only modifies the appearances of phenomena).

\medskip

The standard way out from the problem of 3-space is to choose the
Euclidean 3-space of an inertial frame centered on an inertial
observer and then use the kinematical Poincare' group to connect
different inertial frames. In the absolute Minkowski space-time
(replacing the Newtonian absolute time and absolute Euclidean
3-space) the Euclidean 3-spaces of the inertial frames centered on
an inertial observer A are identified by means of Einstein
convention for the synchronization of clocks: the inertial observer
A sends a ray of light at $x^o_i$ towards the (in general
accelerated) observer B; the ray is reflected towards A at a point P
of B world-line and then reabsorbed by A at $x^o_f$; by convention P
is synchronous with the mid-point between emission and absorption on
A's world-line, i.e. $x^o_P = x^o_i + {1\over 2}\, (x^o_f - x^o_i) =
{1\over 2}\, (x^o_i + x^o_f)$.

\medskip

However, this is not possible in general relativity (GR), where
there is no absolute notion since also space-time becomes dynamical
(with the metric structure satisfying Einstein's equations). The
equivalence principle implies the absence of global inertial frames:
in the restricted class of globally hyperbolic, asymptotically
Minkowskian at spatial infinity, space-times the best we can have
are global non-inertial frames connected by 4-diffeomorphisms (the
gauge group of GR). As a consequence, also in SR we have to face the
problem of reformulating physics in non-inertial frames centered on
accelerated observers \cite{6} as a first step before facing GR
\footnote{Regarding the problem of clock synchronization in presence
of gravity near the Earth let us underline the relevance of the ACES
mission of ESA \cite{5}, programmed for 2013. It will make possible
a measurement of the gravitational redshift of the Earth from the
two-way link among a microwave clock (PHARAO) on the Space Station
and similar clocks on the ground: the proposed microwave link should
make possible the control of effects on the scale of 5 picoseconds.
This will be a test of post-Newtonian gravity in the framework of
Einstein's geometrical view of gravitation: the redshift is a
measure of the $1/c^2$ deviation of post-Newtonian null geodesics
from Minkowski ones. This is a non-perturbative effect (requiring a
re-summation of the whole perturbative expansion) for every quantum
field theory, which has to fix the background (and therefore the
associated light-cones) to be able to define the quantum Fock space
of the theory. As a consequence soon we will also need a formulation
of relativistic entanglement in non-inertial frames in presence of
post-Newtonian gravity.}.

\bigskip

A metrology-oriented description of non-inertial frames  in SR has
been  done with the {\it 3+1 point of view} and the use of
observer-dependent Lorentz scalar radar 4-coordinates \cite{6,7}.
Let us give the world-line $x^{\mu}(\tau)$ of an arbitrary time-like
observer carrying a standard atomic clock: $\tau$ is an arbitrary
monotonically increasing function of the proper time of this clock.
Then we give an admissible 3+1 splitting of Minkowski space-time,
namely a nice foliation with space-like instantaneous 3-spaces
$\Sigma_{\tau}$: it is the mathematical idealization of a protocol
for clock synchronization (all the clocks in the points of
$\Sigma_{\tau}$ sign the same time of the atomic clock of the
observer). On each 3-space $\Sigma_{\tau}$ we choose curvilinear
3-coordinates $\sigma^r$ having the observer as origin. These are
the radar 4-coordinates $\sigma^A = (\tau; \sigma^r)$. If $x^{\mu}
\mapsto \sigma^A(x)$ is the coordinate transformation from the
Cartesian 4-coordinates $x^{\mu}$ of a reference inertial observer
to radar coordinates, its inverse $\sigma^A \mapsto x^{\mu} =
z^{\mu}(\tau ,\sigma^r)$ defines the {\it embedding} functions
$z^{\mu}(\tau ,\sigma^r)$ describing the 3-spaces $\Sigma_{\tau}$ as
embedded 3-manifold into Minkowski space-time. The induced 4-metric
on $\Sigma_{\tau}$ is the following functional of the embedding
${}^4g_{AB}(\tau ,\sigma^r) = [z^{\mu}_A\, \eta_{\mu\nu}\,
z^{\nu}_B](\tau ,\sigma^r)$, where $z^{\mu}_A = \partial\,
z^{\mu}/\partial\, \sigma^A$ and ${}^4\eta_{\mu\nu} = \sgn\, (+---)$
is the flat metric ($\sgn = \pm 1$ according to either the particle
physics $\sgn = 1$ or the general relativity $\sgn = - 1$
convention). While the 4-vectors $z^{\mu}_r(\tau ,\sigma^u)$ are
tangent to $\Sigma_{\tau}$, so that the unit normal $l^{\mu}(\tau
,\sigma^u)$ is proportional to $\epsilon^{\mu}{}_{\alpha
\beta\gamma}\, [z^{\alpha}_1\, z^{\beta}_2\, z^{\gamma}_3](\tau
,\sigma^u)$, we have $z^{\mu}_{\tau}(\tau ,\sigma^r) = [N\, l^{\mu}
+ N^r\, z^{\mu}_r](\tau ,\sigma^r)$ ($N(\tau ,\sigma^r) = \sgn\,
[z^{\mu}_{\tau}\, l_{\mu}](\tau ,\sigma^r)$ and $N_r(\tau ,\sigma^r)
= - \sgn\, g_{\tau r}(\tau ,\sigma^r)$ are the lapse and shift
functions).\medskip

The foliation is nice and admissible if it satisfies the conditions
\footnote{These conditions imply that global {\it rigid} rotations
are forbidden in relativistic theories. In Ref.\cite{6,8} there is
the expression of the admissible embedding corresponding to a 3+1
splitting of Minkowski space-time with parallel space-like
hyper-planes (not equally spaced due to a linear acceleration)
carrying differentially rotating 3-coordinates without the
coordinate singularity of the rotating disk. It is the first
consistent global non-inertial frame of this type.}: \hfill\break
 1) $N(\tau ,\sigma^r) > 0$ in every point of
$\Sigma_{\tau}$ (the 3-spaces never intersect, avoiding the
coordinate singularity of Fermi coordinates);\hfill\break
 2) $\sgn\, {}^4g_{\tau\tau}(\tau ,\sigma^r) > 0$, so to avoid the
 coordinate singularity of the rotating disk, and with the positive-definite 3-metric
${}^3g_{rs}(\tau ,\sigma^u) = - \sgn\, {}^4g_{rs}(\tau ,\sigma^u)$
having three positive eigenvalues (these are the M$\o$ller
conditions \cite{6,8});\hfill\break
 3) all the 3-spaces $\Sigma_{\tau}$ must tend to the same space-like
 hyper-plane at spatial infinity (so that there are always asymptotic inertial
observers to be identified with the fixed stars).\medskip

In the 3+1 point of view the 4-metric ${}^4g_{AB}(\tau ,\vec \sigma
)$ on $\Sigma_{\tau}$ has the components $\sgn\, {}^4g_{\tau\tau} =
N^2 - N_r\, N^r$, $- \sgn\, {}^4g_{\tau r} = N_r = {}^3g_{rs}\,
N^s$, ${}^3g_{rs} = - \sgn\, {}^4g_{rs} = \sum_{a=1}^3\,
{}^3e_{(a)r}\, {}^3e_{(a)s} = {\tilde \phi}^{2/3}\, \sum_{a=1}^3\,
e^{2\, \sum_{\bar b =1}^2\, \gamma_{\bar ba}\, R_{\bar b}}\,
V_{ra}(\theta^i)\, V_{sa}(\theta^i)$), where ${}^3e_{(a)r}(\tau
,\sigma^u)$ are cotriads on $\Sigma_{\tau}$, ${\tilde \phi}(\tau
,\sigma^r) = \sqrt{det\, {}^3g_{rs}(\tau ,\sigma^r)}$ is the
3-volume element on $\Sigma_{\tau}$, $\lambda_a(\tau ,\sigma^r) =
[{\tilde \phi}^{1/3}\, e^{\sum_{\bar b =1}^2\, \gamma_{\bar ba}\,
R_{\bar b}}](\tau ,\sigma^r)$ are the positive eigenvalues of the
3-metric ($\gamma_{\bar aa}$ are suitable numerical constants) and
$V(\theta^i(\tau ,\sigma^r))$ are diagonalizing rotation matrices
depending on three Euler angles. The components ${}^4g_{AB}$ or the
quantities $N$, $N_r$, $\gamma$, $R_{\bar a}$, $\theta^i$, play the
role of the {\it inertial potentials} generating the relativistic
apparent forces in the non-inertial frame. It can be shown
\cite{6,8} that the Newtonian inertial potentials are hidden in the
functions $N$, $N_r$ and $\theta^i$. The extrinsic curvature
${}^3K_{rs}(\tau, \sigma^u) = [{1\over {2\, N}}\, (N_{r|s} + N_{s|r}
- \partial_{\tau}\, {}^3g_{rs})](\tau, \sigma^u)$, describing the
{\it shape} of the instantaneous 3-spaces of the non-inertial frame
as embedded 3-manifolds of Minkowski space-time, is a secondary
inertial potential functional of the independent inertial potentials
${}^4g_{AB}$.\medskip

The description of isolated systems (particles, strings, fields,
fluids) admitting a Lagrangian formulation in the non-inertial
frames of SR is done by means of {\it parametrized Minkowski
theories} \cite{6,7}. The matter variables are replaced with new
ones knowing the 3-spaces $\Sigma_{\tau}$. For instance a
Klein-Gordon field $\tilde \phi (x)$ will be replaced with
$\phi(\tau ,\sigma^r) = \tilde \phi (z(\tau ,\sigma^r))$; the same
for every other field. Instead for a relativistic particle with
world-line $x^{\mu}(\tau )$ we must make a choice of its energy
sign: then it will be described by 3-coordinates $\eta^r(\tau )$
defined by the intersection of the world-line with $\Sigma_{\tau}$:
$x^{\mu}(\tau ) = z^{\mu}(\tau ,\eta^r(\tau ))$. Differently from
all the previous approaches to relativistic mechanics, the dynamical
configuration variables are the 3-coordinates $\eta^r_i(\tau)$ and
not the world-lines $x^{\mu}_i(\tau)$ (to rebuild them in an
arbitrary frame we need the embedding defining that frame!). Then
the matter Lagrangian is coupled to an external gravitational field
and the external 4-metric is replaced with the 4-metric $g_{AB}(\tau
,\sigma^r)$ of an admissible 3+1 splitting of Minkowski space-time.
With this procedure we get a Lagrangian depending on the given
matter and on the embedding $z^{\mu}(\tau ,\sigma^r)$, which is
invariant under {\it frame-preserving diffeomorphisms} \cite{9}. As
a consequence, there are four first-class constraints (an analogue
of the super-Hamiltonian and super-momentum constraints of canonical
gravity) implying that the embeddings $z^{\mu}(\tau ,\sigma^r)$ are
{\it gauge variables}, so that all the admissible non-inertial or
inertial frames are gauge equivalent, namely physics does {\it not}
depend on the clock synchronization convention and on the choice of
the 3-coordinates $\sigma^r$: only the appearances of phenomena
change by changing the notion of instantaneous 3-space. Even if the
gauge group is formed by the frame-preserving diffeomorphisms, the
matter energy-momentum tensor allows the determination of the ten
conserved Poincare' generators $P^{\mu}$ and $J^{\mu\nu}$ (assumed
finite) of every configuration of the system; they are non-local
quantities knowing the whole 3-space!.

\bigskip

If we restrict ourselves to inertial frames, we can define the {\it
inertial rest-frame instant form of dynamics for isolated systems}
by choosing the 3+1 splitting corresponding to the intrinsic
inertial rest frame of the isolated system centered on an inertial
observer: the instantaneous 3-spaces, named Wigner 3-space due to
the fact that the 3-vectors inside them are Wigner spin-1 3-vectors
\cite{7}, are orthogonal to the conserved 4-momentum $P^{\mu}$ of
the configuration. In Ref.\cite{6} there is the extension to
admissible non-inertial rest frames, where $P^{\mu}$ is orthogonal
to the asymptotic space-like hyper-planes to which the instantaneous
3-spaces tend at spatial infinity. This non-inertial family of 3+1
splittings is the only one admitted by the asymptotically
Minkowskian space-times covered by the canonical gravity formulation
of Refs.\cite{10,11}.
\medskip

In the inertial rest frames we can get the explicit form of the
Poincare' generators (in particular of the Lorentz boosts, which,
differently from the Galilei ones, are interaction dependent).  We
can also give the final solution to the old problem of the
relativistic extension of the Newtonian center of mass of an
isolated system. In its rest frame there are {\it only} three
notions of collective variables, which can be built by using {\it
only} the Poincare' generators (they are {\it non-local} quantities
knowing the whole $\Sigma_{\tau}$) \cite{12}: the canonical
non-covariant Newton-Wigner center of mass (or center of spin), the
non-canonical covariant Fokker-Pryce center of inertia and the
non-canonical non-covariant M$\o$ller center of energy. All of them
tend to the Newtonian center of mass in the non-relativistic limit.
See Ref.\cite{7} for the M$\o$ller non-covariance world-tube around
the Fokker-Pryce 4-vector identified by these collective variables.
As shown in Refs.\cite{12,13,14} these three variables can be
expressed as known functions of the rest time $\tau$, of the
canonically conjugate Jacobi data (frozen Cauchy data) $\vec z =
Mc\, {\vec x}_{NW}(0)$ (${\vec x}_{NW}(\tau )$ is the standard
Newton-Wigner 3-position) and $\vec h = \vec P/Mc$, of the invariant
mass $Mc = \sqrt{\sgn\, P^2}$ of the system and of its rest spin
${\vec {\bar S}}$.  As a consequence, every isolated system (i.e. a
closed universe) can be visualized as a decoupled non-covariant
collective (non-local) pseudo-particle described by the frozen
Jacobi data $\vec z$, $\vec h$ carrying a {\it pole-dipole
structure} (see also Ref.\cite{15}), namely the invariant mass and
the rest spin of the system, and with an associated {\it external}
realization of the Poincare' group. The universal breaking of
Lorentz covariance is connected to this decoupled non-local
collective variable and is irrelevant because all the dynamics of
the isolated system leaves inside the Wigner 3-spaces and is
Wigner-covariant. In each Wigner 3-space $\Sigma_{\tau}$ there is a
{\it unfaithful internal} realization of the Poincare' algebra,
whose generators are built by using the energy-momentum tensor of
the isolated system. While the internal energy and angular momentum
are $Mc$ and ${\vec {\bar S}}$ respectively, the internal 3-momentum
vanishes: it is the rest frame condition. Also the internal Lorentz
boost (whose expression in presence of interactions is given for the
first time) vanishes: this condition identifies the covariant
non-canonical Fokker-Pryce center of inertia as the natural inertial
observer origin of the 3-coordinates $\sigma^r$ in each Wigner
3-space. As a consequence \cite{16} there are three pairs of second
class (interaction-dependent) constraints eliminating the internal
3-center of mass and its conjugate momentum inside the Wigner
3-spaces \cite{16}: this avoids a double counting of the collective
variables and allows to re-express the dynamics only in terms of
internal Wigner-covariant relative variables. As a consequence, we
find that disregarding the unobservable center of mass all the
dynamics is described only by relative variables: this is a form of
{\it weak relationism} without the heavy foundational problem of
approaches like the one in Ref.\cite{17}.
\medskip

In the case of relativistic particles the reconstruction of their
world-lines requires a complex interaction-dependent procedure
delineated in Ref.\cite{14}. The final derived world-lines
$x^{\mu}_i(\tau)$ turn out to be non-canonical predictive
coordinates, i.e. already at the classical level there is a
non-commutative structure implied by Lorentz signature. See
Ref.\cite{16} for the comparison with the other formulations of
relativistic mechanics developed for the study of the problem of
{\it relativistic bound states}. In Refs.\cite{14,18} there is the
explicit form of the Lorentz boosts for some interacting systems.

\bigskip

In this framework it has been possible to obtain a relativistic
formulation of the classical background of atomic physics,
considered as an effective theory of positive-energy scalar (or
spinning) particles with mutual Coulomb interaction plus the
transverse electro-magnetic field of the radiation gauge valid for
energies below the threshold of pair production. As shown in
Refs.\cite{13} and \cite{16}, this has been possible by considering
Grassmann-valued electric charges for the particles ($Q_i^2 = 0$,
$Q_i\, Q_j = Q_j\, Q_i \not= 0$ for $i \not= j$). It allows a) to
make an ultraviolet regularization of Coulomb self-energies; b) to
make an infrared regularization eliminating the photon emission; c)
to express the Lienard-Wiechert potentials only in terms of the
3-coordinates $\eta^r_i(\tau )$ and the conjugate 3-momenta
$\kappa_{ir}(\tau)$ in a way independent from the used (retarded,
advanced,..) Green function. All this amount to reformulate the
dynamics of the one-photon exchange as a Cauchy problem with well
defined potentials. Moreover there is a canonical transformation
\cite{16} sending the above system in a transverse radiation field
(in- or out-fields) decoupled, in the global rest frame, from
Coulomb-dressed particles with a mutual interaction described by the
sum of the Coulomb potential plus the Darwin potential. Therefore
for the first time we are able to obtain results, previously derived
from instantaneous approximations to the Bethe-Salpeter equation for
the description of relativistic bound states (see the bibliography
of Ref.\cite{13}), starting from the classical theory. Moreover, for
the first time, at least at the classical level, we have been able
to avoid the Haag theorem according to which the interaction picture
does not exist in QFT.

\bigskip

In refs.\cite{21} there is the {\it multi-temporal quantization} of
positive-energy free scalar and spinning particles in a family of
global non-inertial frames of Minkowski space-time with space-like
hyper-planes as 3-spaces and differentially rotating 3-coordinates
defined in Ref.\cite{6}. We take the point of view {\it not to
quantize the inertial effects} (the appearances of phenomena): the
embedding $z^{\mu}(\tau ,\sigma^r)$ remains a c-number and we get
results compatible with atomic spectra. Instead the problem of the
reformulation of particle physics in non-inertial frames is unsolved
due to the no-go theorem of Ref.\cite{22} showing the existence of
obstructions to the unitary evolution of a massive Klein-Gordon
field between two space-like surfaces of Minkowski space-time. This
problem has to be reformulated as the search of the class of
admissible 3+1 splittings of Minkowski space-time admitting unitary
evolution after quantization: this would allow to check whether the
hypothesis of non-quantized inertial effects is valid also in field
theory (it will be a crucial point for quantum gravity!).

\bigskip

The previous framework allowed to find a new formulation of {\it
relativistic quantum mechanics and entanglement}, which is developed
in Ref. \cite{19}.
\medskip

As already said, in Galilei space-time non-relativistic QM, where
all the main results about entanglement are formulated, describes a
composite system with two (or more) subsystems with a Hilbert space
which is the tensor product of the Hilbert spaces of the subsystems:
$H = H_1 \otimes H_2$. This type of spatial separability is named
{\it the zeroth postulate} of quantum mechanics. However, when the
two subsystems are mutually interacting, one makes a unitary
transformation to the tensor product of the Hilbert space $H_{com}$
describing the decoupled Newtonian center of mass of the two
subsystems and of the Hilbert space $H_{rel}$ of relative variables:
$H = H_1 \otimes H_2 = H_{com} \otimes H_{rel}$. This allows to use
the method of separation of variables to split the Schroedinger
equation in two equations: one for the free motion of the center of
mass and another, containing the interactions, for the relative
variables (this equation describes both the bound and scattering
states). A final unitary transformation of the Hamilton-Jacobi type
allows to replace $H_{com}$ with $H_{com, HJ}$, the Hilbert space in
which the decoupled center of mass is frozen and described by
non-evolving Jacobi data. Therefore we have $H = H_1 \otimes H_2 =
H_{com} \otimes H_{rel} = H_{com, HJ} \otimes H_{rel}$.\medskip

While at the non-relativistic level these three descriptions are
unitary equivalent, this no more true in relativistic quantum
mechanics, due to the previously described problems arising from
Lorentz signature like the need of clock synchronization and the
non-covariance of the (non-local) canonical center of mass.
\medskip

In the approach of Ref.\cite{19} we quantize the frozen Jacobi data
of the canonical non-covariant decoupled center of mass and the
Wigner-covariant relative variables on the Wigner hyper-plane. Since
the center of mass is decoupled, its non-covariance is irrelevant:
like for the wave function of the universe, who will observe it?
Moreover its non-local nature implies that it is not a locally
measurable quantity. Finally the use of the static Jacobi data for
the external center of mass avoids the causality problems connected
with the instantaneous spreading of wave packets (the Hegerfeldt
theorem \cite{20}). This viewpoint is in accord with relativistic
bound states and relativistic atomic physics
\medskip

The need of clock synchronization for the definition of the
instantaneous 3-spaces and the non-local and non-covariant
properties of the decoupled relativistic center of mass, described
by the frozen Jacobi data $\vec z$ and $\vec h$, imply that the only
consistent relativistic quantization is based on the Hilbert space
$H = H_{com, HJ} \otimes H_{rel}$ ($H_{com, HJ}$ is the Hilbert
space of the external center of mass in the Hamilton-Jacobi
formulation, while $H_{rel}$ is the Hilbert space of the internal
relative variables). The Hilbert space $H$  is not unitarily
equivalent to $H_1 \otimes H_2 \otimes ...$, where $H_i$ are the
Hilbert spaces of the individual particles. This is due to the fact
that already in the non-interacting two-particle case, in the tensor
product of two quantum Klein-Gordon fields, $\phi_1(x_1)$ and
$\phi_2(x_2)$, most of the states correspond to configurations in
Minkowski space-time in which one particle may be present in the
absolute future of the other particle, because the two times $x^o_1$
and $x^o_2$ are totally uncorrelated, or in other words there is no
notion of instantaneous 3-space (clock synchronization convention).
Also the scalar products in the two formulations are completely
different as shown in Ref.\cite{23}. In S-matrix theory this problem
is eliminated by avoiding the interpolating states at finite (the
problem of the Haag theorem) and going the the asymptotic (in the
times $x^o_i$) limit of the free in- and out- states. However in
atomic physics we need interpolating states, and not S-matrix, to
describe a laser beam resonating in a cavity and intersected by a
beam of atoms!\medskip

We have also $H \not= H_{com} \otimes H_{rel}$, because if instead
of $\vec z = Mc\, {\vec x}_{NW}(0)$ we use the evolving (non-local
and non-covariant) Newton-Wigner position operator ${\vec
x}_{NW}(\tau )$, then we get a violation of relativistic causality
because the center-of-mass wave packets spread instantaneously as
shown by the Hegerfeldt theorem \cite{20}.\medskip

Therefore the only consistent Hilbert space is $H = H_{com, HJ}
\otimes H_{rel}$, whose non-relativistic limit is the corresponding
Newtonian Hilbert space. The main complication is the definition of
$H_{rel}$, because we must take into account the three pairs of
(interaction-dependent) second-class constraints eliminating the
internal 3-center of mass inside the Wigner 3-spaces. When we are
not able to make the elimination at the classical level and
formulate the dynamics only in terms of Wigner-covariant relative
variables, we have to quantize the particle Wigner-covariant
3-variables $\eta^r_i$, $\kappa_{ir}$ and then to define the
physical Hilbert space by adding the quantum version of the
constraints a la Gupta-Bleuler.

\bigskip

The quantization defined in Ref.\cite{19} leads to a first
formulation of a theory for {\it relativistic entanglement}. The non
validity of the zeroth postulate and the {\it non-locality} of
Poincare' generators imply a {\it kinematical non-locality} and a
{\it kinematical spatial non-separability} introduced by special
relativity, which reduce the relevance of {\it quantum non-locality}
in the study of the foundational problems of quantum mechanics which
have to be rephrased in terms of relative variables. Einstein's
notion of separability is not valid since in $H = H_{com, HJ}
\otimes H_{rel}$ the composite system must be described by means of
relative variables in a Wigner 3-space (this is a type of weak form
of relationism different from the formulations induced by the Mach
principle like in Ref.\cite{17}).
\medskip

The relativistic formulation of problems like the relevance of
decoherence \cite{24} for the selection of preferred robust pointer
bases and the emergence of quasi-classical macroscopic objects from
quantum constituents will have to be done in terms of relative
variables. Moreover, the control  of Poincare' kinematics will force
to reformulate the experiments connected with Bell inequalities and
teleportation in terms of isolated systems containing: a) the
observers with their measuring apparatus (Alice and Bob as
macroscopic quasi-classical objects); b) the particles of the
protocol (but now the ray of light, the "photons" carrying the
polarization, move along null geodesics); c) the environment
(macroscopic either quantum or quasi-classical object). All these
problems disappear as $1/c$ corrections in experiments on Earth, but
will be relevant for space physics. A preliminary step for
understanding this framework would be to reformulate the
non-relativistic theory of entanglement in the Hilbert space $H =
H_{com, HJ} \otimes H_{rel}$ and to forget about the decoupled
center-of-mass Hilbert space $H_{com, HJ}$.

\bigskip

The final challenge will be a consistent inclusion of the
gravitational field, at least at the post-Newtonian level! See
Refs.\cite{11} for the use of 3+1 splittings and of radar
4-coordinates in ADM canonical tetrad gravity in globally
hyperbolic, asymptotically Minkowskian at spatial infinity,
space-times. As shown in Ref.\cite{25}, the dynamical nature of
space-time implies that each solution of Einstein's equations
dynamically selects a preferred 3+1 splitting of the space-time,
namely in GR the instantaneous 3-spaces  are dynamically determined
except for the trace of the extrinsic curvature (the York time
${}^3K(\tau, \sigma^r)$): this inertial gauge variable is the
general relativistic remnant of the special relativistic gauge
freedom in clock synchronization.

\end{document}